\documentclass[pre, aps,twocolumn,superscriptaddress,longbibliography,floatfix,notitlepage]{revtex4-1}

\usepackage[utf8x]{inputenc}
\usepackage[english]{babel}
\usepackage[T1]{fontenc}
\usepackage{lmodern}

\usepackage{dsfont}
\usepackage{amsmath,amsfonts,amssymb,amsthm,bm,times}
\usepackage[colorlinks={true}, citecolor={blue}, filecolor={blue}, linkcolor={blue}, urlcolor={blue}]{hyperref}
\usepackage{graphicx,color}
\usepackage[caption=false]{subfig}
\usepackage{microtype}
\usepackage{braket}

\newcommand\abs[1]{\left|#1\right|}

\begin{document}
	
	\title{Quantum Optical Two-Atom Thermal Diode}
	
	\author{Cahit Kargi}
	\affiliation{Department of Physics, Ko\c{c} University, 34450 Sariyer, Istanbul TURKEY}
	
	\author{M. Tahir Naseem}
	\affiliation{Department of Physics, Ko\c{c} University, 34450 Sariyer, Istanbul TURKEY}
	
	\author{Tom\'{a}\v{s} Opatrn\'{y}}
	\affiliation{Department of Optics, Palack\'{y} University, 17. listopadu 50, 77146 Olomouc, 
		Czech Republic}
	
	\author{\"{O}zg\"{u}r E. M\"{u}stecapl{\i}o\u{g}lu}
	\email{omustecap@ku.edu.tr}
	\affiliation{Department of Physics, Ko\c{c} University, 34450 Sariyer, Istanbul TURKEY}
	
	\author{Gershon Kurizki}
	\affiliation{Department of Chemical and Biological Physics, Weizmann Institute of Science, 
		Rehovot 7610001, Israel}
	
	\date{\today}
	
	\begin{abstract}
		We put forward a quantum-optical model for a thermal diode based on heat transfer between two thermal baths through a pair of interacting qubits. We find that if the qubits are coupled by a Raman field that induces an anisotropic interaction, heat flow can become non-reciprocal and undergoes rectification even if the baths have 
		equal
		dissipation rates and/or the qubits are resonant. 
		The heat flow rectification is explained by four-wave mixing and Raman transitions between dressed states of the interacting qubits and is governed by a
		global master equation. 
		The anisotropic two-qubit interaction is the key for this present simple quantum thermal diode, 
		whose resonant operation allows for high-efficiency rectification of large heat currents. Effects of spatial
		overlap of the baths are addressed. We also discuss the possible realizations of the model system in various platforms including optomechanical systems, systems of trapped ions, and circuit QED.
	\end{abstract}

	\maketitle
	
	
	\section{\label{sec:Intro}Introduction}
	
	
	A heat diode (HD) is a device that conducts heat under thermal bias in the direction chosen as forward – say, from a heat bath on the left side to a cold bath on the right side of the device – but insulates heat flow under the reverse (backward) thermal bias, i.e., from a hot bath on the right to a cold bath on the left, according to the choice made above~\cite{RevModPhys.78.217,ROBERTS2011648,RevModPhys.84.1045,BENENTI20171}. The HD proposals and experimental realizations in solid-state~\cite{PhysRevLett.88.094302,PhysRevLett.93.184301,PhysRevB.74.214305,PhysRevB.75.214302,PhysRevB.76.020301,doi:10.1143/JPSJ.77.054402}, mesoscopic~\cite{Chang1121,doi:10.1063/1.2191730,Nature460,PhysRevApplied.6.054003}  and quantum systems~\cite{PhysRevLett.107.173902,2058-9565-2-4-044007,arXiv:1801.09312,PhysRevLett.120.060601,PhysRevLett.100.155902,PhysRevLett.116.200601,PhysRevE.90.042142, PhysRevE.89.062109,PhysRevE.94.042135,PhysRevE.95.022128} attest to the keen interest in this subject, motivated by the expectation that HD would become to phononics~\cite{harvester1,harvester2,infoProc,PhysRevApplied.4.014011,heatengine1,heatengine2}, primarily at the nanoscale and quantum domains, what a semiconducting diode is to micro- or nano-electronics~\cite{kouwenhoven_quantized_1991,moldoveanu_nonadiabatic_2007,gudmundsson_time-dependent_2012}. For the realization of these prospects it is essential to acquire a deep understanding of HD operation principles in the quantum domain. Here we set out to resolve the basic issues of  quantum HD operation:

1)      Since HD operation requires reciprocity breaking between forward and backward heat flow at the quantum junction connecting two baths, can we identify a genuine quantum mechanism of such reciprocity breaking? Classically or quasi-classically, an HD junction is commonly viewed as a dissipative ratchet~\cite{harvester1,PhysRevLett.79.10} where rectification is achieved by the left-right asymmetry (tilt) of its energy spectrum, obstructing the heat flow to the “wrong” side. A ratchet is a spatially-extended structure, e.g., a spin chain~\cite{doi:10.1063/1.2828737,Casati1}, but is there an alternative quantum HD model in the case of a single-atom (dot) junction or a junction comprised of two closely-spaced atoms or dots?

2)      A similar question concerns a classical HD mechanism whereby the junction is asymmetrically coupled to the left- and right-heat channels (baths) ~\cite{dames_solid-state_2009, maznev_reciprocity_2013, jezowski_heat_1978} which again requires an extended structure such that the couplings to the two baths are distinct, which may be impossible on nano- or micro-scales. Can there be an effectively asymmetric coupling to the two baths in the quantum domain, notwithstanding their close proximity or overlap?

3)      Once an HD junction has been constructed, can it be controlled, so as to adapt it to the situation at hand?

In this paper we provide affirmative answers to all three basic questions raised above, by putting forward a simple quantum HD scheme:

a)      The proposed scheme is based on two anisotropically interacting qubits. The mechanism responsible for  HD  operation in this scheme is  the left-right asymmetric interaction of the two qubits. It may arise, for example, when an external classical field is aligned with the z-axis of the Bloch sphere of one qubit and with the x-axis of its counterpart, as proposed by Rao and Kurizki~\cite{bhaktavatsala_rao_zeno_2011}. It is universally adaptable to any material, qubit level spacing and temperature and entirely relies on quantum optical tools, i.e. Raman and four-wave mixing transitions that bypass the ratchet (spatial tilt) requirements. Natural realizations can be found in ferromagnetic systems~\cite{moriya_anisotropic_1960,dzyaloshinsky_thermodynamic_1958} or in nuclear spin environments~\cite{tupitsyn_effective_1997}. It can be engineered, as we shall discuss in Sec.~\ref{sec:implementations}, using related model Hamiltonians of various systems such as cavity QED~\cite{knight_quantum_1986,schoendorff_analytic_1990,Phoenix:90},
trapped ions~\cite{PhysRevA.65.032310}, circuit QED~\cite{johansson_optomechanical-like_2014,blais_cavity_2004,blais_quantum-information_2007,zheng_arbitrary_2010}, and optomechanics~\cite{PhysRevA.92.062114,law_effective_1994,law_interaction_1995,Moqadam}.

Previously, for a junction comprised of two interacting qubits, it has been concluded that either the qubits must be off-resonant with each other~\cite{PhysRevE.95.022128} or that the two baths and the qubit-bath couplings need have asymmetry, in addition to the different bath temperatures~\cite{PhysRevE.89.062109,PhysRevE.94.042135,PhysRevE.95.022128}. Here we show that neither of the above  mechanisms is compulsory for our quantum HD. A system of two  anisotropically interacting resonant qubits that are symmetrically coupled to two identical baths (that differ only in temperature) may allow for high-efficiency heat rectification and much higher heat flow than previously suggested schemes.

b)      Our analysis elucidates the requirements that the left- and right-baths be distinct, so that they can be assigned different temperatures and allow for heat flow from one bath to the other, even in case of close proximity or overlap  of the baths. Such a scenario requires an analysis based on a global master equation for the entire setup, as previously stressed~\cite{breuer2002,1367-2630-12-11-113032,NJP-Hofer,0295} rather than on (generally inadequate) local master equations for each of the qubits~\cite{Condmatter-Manrique,OpenSys-Correa}.

c)       An important aspect of our proposed scheme is its controllability through tuning the strength of the two-qubit Raman-coupling field, which can in turn strongly inhibit or permit four-wave mixing and Raman transitions between the system levels via thermal quanta. Thus, the control Raman field acts as the valve that obstructs or enables global heat transport through the junction.

The
paper is organized as follows. In Sec.~\ref{sec:Model}, we describe our model and discuss the physical mechanism behind the diode operation. In Sec.~\ref{sec:HC}, we present
its results for the heat currents and the rectification factors. We discuss possible implementations of this model scheme in Sec.~\ref{sec:implementations}. Finally, we conclude in Sec.~\ref{sec:Conclusion}.
	\section{\label{sec:Model}Model and Physical Mechanism}
\begin{figure}[t]
		\centering
		\includegraphics[width=0.5\textwidth]{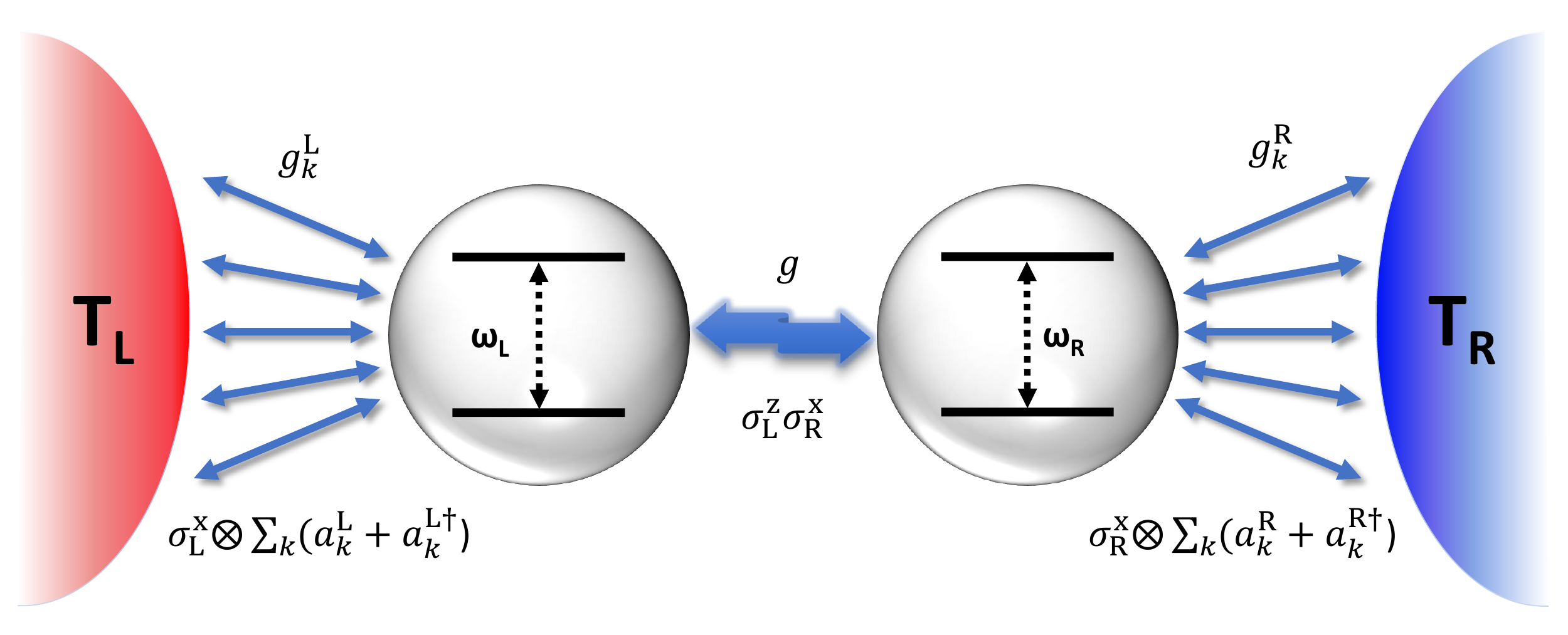}
		\caption{(Color online) Schematic diagram of the quantum thermal diode based on two qubits coupled via anisotropic (Raman-induced) spin-spin interaction $\hat{\sigma}_{\text{L}}^{z}\hat{\sigma}_{\text{R}}^{x}$. The left qubit has transition frequency $\omega_L$ while right qubit has transition frequency $\omega_R$ and the coupling strength between the qubits is denoted by $g$. Each qubit is coupled with a thermal bath and we assume these baths are independent and may have any distinct non-negative temperatures.}
		\label{fig:BareQubits}
	\end{figure}

	We consider a system of two interacting qubits, with transition frequencies $\omega_{\text{L}}$ and $\omega_{\text{R}}$ as shown in Fig.~\ref{fig:BareQubits}. The Hamiltonian of the system is (we take $\hbar = 1$)
	\begin{equation}\label{eq:HamiltonianBare}
	\hat{H} = \frac{\omega_{\text{L}}}{2}\hat{\sigma}_{\text{L}}^{z} + \frac{\omega_{\text{R}}}{2}\hat{\sigma}_{\text{R}}^{z} + g\hat{\sigma}_{\text{L}}^{z}\hat{\sigma}_{\text{R}}^{x},
	\end{equation}
	where $g$ is the coupling strength of the Raman-induced anisotropic exchange interaction between the left (L) and right (R) qubits, and $\hat{\sigma}_\alpha^\beta$
	with $\alpha=\text{L,R}$, $\beta=z,x$ are the Pauli matrices.
	
	In order to derive the global Markovian master equation, we first diagonalize the coupled-qubit system Hamiltonian by a unitary transformation (see Appendix~\ref{AppendixME}) that yields the dressed-system Hamiltonian in the form
	\begin{equation}\label{eq:diaghamil}
	\hat{\tilde{H}} = \frac{\omega_{\text{L}}}{2}\hat{\tilde{\sigma}}_{\text{L}}^{z} + \frac{\Omega}{2}\hat{\tilde{\sigma}}_{\text{R}}^{z},
	\end{equation}
	where $\Omega = \sqrt{\omega_{\text{R}}^{2} + 4g^{2}}$. The transformed Pauli matrices are denoted by $\hat{\tilde{\sigma}}_\alpha^\beta$. 
	The master equation is derived in Appendix~\ref{AppendixME}. In 
	the
	interaction picture it has the form
	\begin{eqnarray}\label{eq:master}
	&\dot{\hat{\rho}}& = \hat{\mathcal{L}}_{\text{LL}} +\hat{ \mathcal{L}}_{\text{RR}},
	\end{eqnarray}
	where $\hat{ \mathcal{L}}_{\text{LL}}$ and $\hat{ \mathcal{L}}_{\text{RR}} $ are Liouville 
	superoperators that describe energy exchange of the system with the baths, whose spectral response functions are given in Appendix~\ref{AppendixME}. 
	 In the derivation of Eq.~(\ref{eq:master}), we have assumed that the baths are independent and each bath is physically connected with its corresponding qubit. Since this can be difficult to implement in cases of proximity between the qubits, we discuss in Appendix~\ref{AppendixB} the case where each bath is physically connected with both qubits.
	 
\begin{figure}[t]
		\centering
		\includegraphics[width=0.5\textwidth]{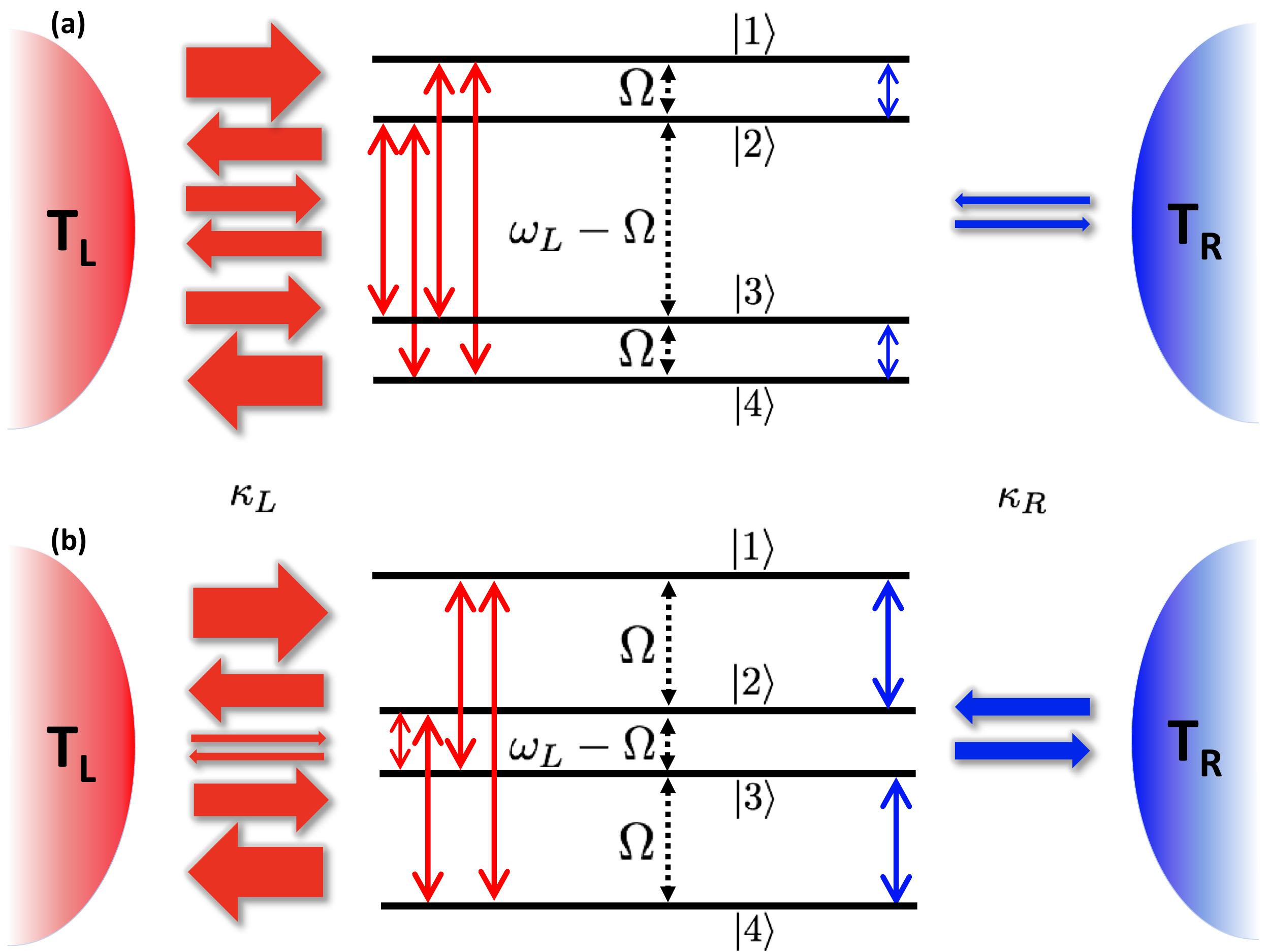}
		\caption{(Color online) Same systeme, but with the system Hamiltonian represented by a diagonalized form. The left bath induces transitions $1 \leftrightarrow 3$, $2 \leftrightarrow 3$, $1 \leftrightarrow 4$ and $2 \leftrightarrow 4$, whereas the right bath  induces transitions $1 \leftrightarrow 2$ and $3 \leftrightarrow 4$.
			(a) Large coupling $g$, (b) small coupling $g$ for resonant qubits.
			In both cases the left bath is at a higher temperature than the right bath.}
		\label{fig:dressedLevels}
	\end{figure}

	 Although Eq.~(\ref{eq:master}) appears to describe two disconnected L and R subsystems it, in fact, allows for excitation exchange between these subsystems, as required for a heat diode (HD) (see Appendix~\ref{app:dynamics}). The structure
	of the dissipators in Eqs.~(\ref{eq:L_L}) and (\ref{eq:L_R}) allows us to identify a simple heat valve mechanism in the heat transport; 
	$\hat{ \mathcal{L}}_{\text{RR}}$ as well as the first two terms in $\hat{ \mathcal{L}}_{\text{LL}}$ represent the local heat transport channels that couple the baths with the corresponding dressed states  $\ket{i}$ of the qubits, but do not contribute to HD operation. The last four terms in Eq.~(\ref{eq:L_R}) describe the heat transport through the global channels between the two baths, which are the only channels that matter for HD. The different channels are characterized by frequencies $\omega_{ij}$ in Fig.~\ref{fig:dressedLevels} with $i,j=1,2,3,4$; $\omega_{13} = \omega_{24} = \omega_{\text{L}}$, $\omega_{14} = \omega_{\text{L}} + \Omega$, $\omega_{23}=\omega_L-\Omega$ and $\omega_{12} = \omega_{34} = \Omega$. 
The channel with spectral response function at frequency $\pm\omega_{23} = \pm(\omega_{\text{L}} -\Omega)$
	transfer the heat via single excitation exchange (flip-flop) through the qubits; while the one at frequency 
	$\pm\omega_{14} = \pm(\omega_{\text{L}} + \Omega)$ 
	transfers the heat via a two quanta (Raman) process. In order to open one or both of the global heat transfer channels, which are the only ones that contribute to HD operation, the temperature of the left bath $T_\text{L}$ needs to be sufficiently high.
As the transition $|4\rangle \to |1\rangle $ requires high energy quanta, $T_{\text{L}}$ may not 
be sufficient to transfer heat by this two-quanta process. As a result, the corresponding channel can be completely or partially obstructed for large  $g$. 
	Therefore,
	when $T_{\text{L}}>\omega_{\text{L}} + \Omega>T_{\text{R}}$ heat flows from left to right through both local and global channels,
	but for
	$T_{\text{R}} >T_{\text{L}}$ 
	the double excitation channel is obstructed and the heat flow decreases.

	\begin{figure}[t]
		\centering
		\includegraphics[width=0.5\textwidth]{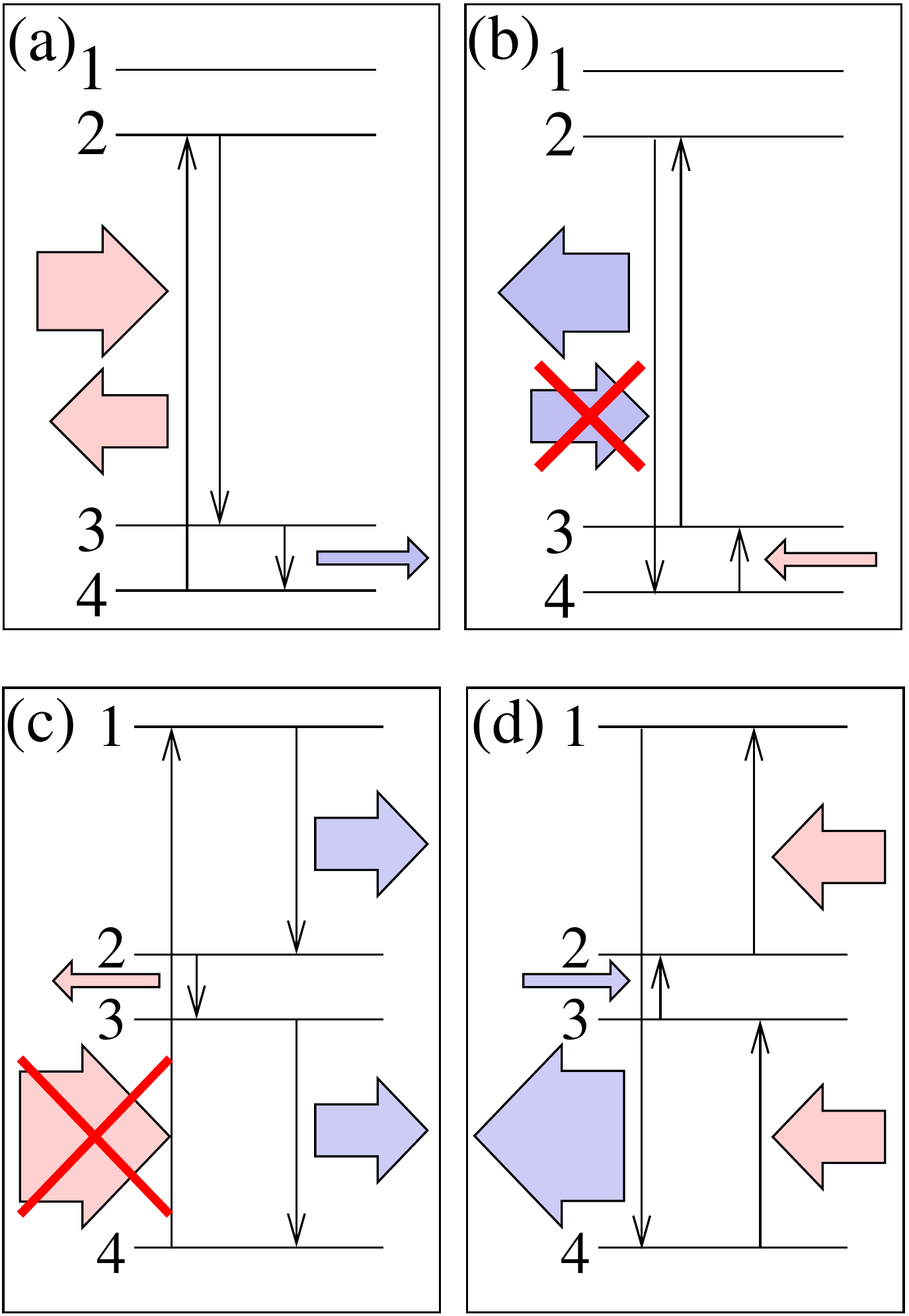}
		\caption{(Color online) Examples of processes that rectify   heat current. Panels (a) and (b) correspond to small $g$,  panels (c) and (d) to large $g$; panels (a) and (c) show the left bath is hotter than the right one, and in panels (b) and (d) it is vice versa. For small $g$, heat can flow from left to right, e.g., via the Raman cycle  
			(4234)
			(a), whereas the opposite cycle 
			(4324)
			(b) is inhibited because the cold reservoir cannot excite the $|3\rangle \to |2\rangle $ transition. For large $g$ and  for  temperatures of the hot reservoir that can not excite $|4\rangle \to |1\rangle $ transition, the four-wave mixing cycle 
			(41234)
			(c) may not be achieved , whereas the opposite cycle  
			(43214)
			(d) can conduct heat from right to left.}
		\label{fig:mechanism}
	\end{figure} 
	
	The rectification effect can 
	be explained by considering 
	possible 
	cycles between the four states $|1\rangle, |2\rangle, |3\rangle, |4\rangle$ 
	which transmit heat from the hot to the cold reservoir via global channels. In principle, there are three possible Raman cycles 
	(3213), (4214), (4234),  and their inverses [here (3213) means the sequence of transitions $|3\rangle \to |2\rangle \to |1\rangle \to |3\rangle$, etc.].
	In addition, three four-wave mixing cycles  
	(41234), (43124), (41324), 
	and their inverses are also possible. Among the four-wave mixing cycles, only the global cycle (41234) transfers heat between the reservoirs while the remaining two keep the energies of the reservoirs unchanged. Depending on the bath temperatures, some cycles can be inhibited in one direction because the reservoir cannot provide photons of sufficient energy to excite particular transitions. As an example, consider 
	the Raman cycle (4234)
	for small $g$ as shown in Figs.~\ref{fig:mechanism}(a) and \ref{fig:mechanism}(b). If the left bath is hot, i.e., $T_{\text{L}} >T_{\text{R}}$ (Fig.~\ref{fig:mechanism}(a)) the cycle may be completed. 
	However, if $T_{\text{R}} >T_{\text{L}}$  (Fig.~\ref{fig:mechanism}(b)) then left bath can not excite the $|3\rangle \to |2\rangle$ transition, consequently the heat transfer from right to left is inhibited and the device rectifies heat from left to right. 
	
	As another example, consider the four-wave mixing cycle 
	(41234)
	for large  $g$ as shown in  Figs.~\ref{fig:mechanism}(c) and \ref{fig:mechanism}(d). If the left bath is hot (Fig.~\ref{fig:mechanism}(c)) but not hot enough to excite the $|4\rangle \to |1\rangle$ transition, 
	the heat transfer from the left to the right bath is inhibited. On the contrary, if the right bath is hot (Fig.~\ref{fig:mechanism}(d)) the reverse cycle may be completed without obstruction so that the device rectifies from right to left. 
	
	The proposed mechanism of the non-reciprocal heat transport relies neither on the frequency difference of the qubits nor on the asymmetry of the dissipation rates of the baths. In our model high rectification 
	can be obtained even for resonant qubits with symmetric couplings to their reservoirs. 
	This is in contrast to other diode models that rely on asymmetric qubit-bath couplings and difference in qubit frequencies to attain rectification ~\cite{PhysRevE.94.042135, PhysRevE.95.022128}.

	\section{\label{sec:HC}Heat Currents and Rectification Factor}
	In order to characterize the heat flow in our system, we calculate the heat currents  $\mathcal{J}_{\text{R}}$ 
	and $\mathcal{J}_{\text{L}}$ 
	from the two baths. 
	According to the definition~\cite{heatCurrent} we have
	\begin{equation}\label{eq:heatcurrent}
	\mathcal{J}_{\text{R}} = \text{Tr}[\hat{ \mathcal{L}}_{\text{R}}\hat{\tilde{H}}].
	\end{equation}
	The first law of thermodynamics 
	requires
	that at steady state the heat currents 
	satisfy  $\mathcal{J}_{\text{R}}=-\mathcal{J}_{\text{L}} $ 
	(a positive value of the
	heat current indicates heat flowing from 
	the
	bath into the system). 
	Here, we only report $\mathcal{J}_{\text{R}}$ and relation for $\mathcal{J}_{\text{L}}$ is given in Appendix~\ref{AppendixC}. 
	Using Eq.~(\ref{eq:master}) in Eq.~(\ref{eq:heatcurrent}) the heat current 
	from the
	right bath is
	\begin{equation}\label{rightHC}
	\mathcal{J}_{\text{R}} = -\frac{1}{2}\kappa_{\text{R}}\Omega\cos^2\theta[1 + (2\bar{n}_\text{R} +1)(\cos\theta\langle\hat{\sigma}_{\text{R}}^{z}\rangle + \sin\theta\langle\hat{\sigma}_{\text{L}}^{z}\hat{\sigma}_{\text{R}}^{x}\rangle)].
	\end{equation}
	Out of the  heat currents one can
	calculate the rectification factor~\cite{RevModPhys.84.1045}, 
	\begin{equation}\label{measure}
	\mathcal{R} = \frac{\abs{\mathcal{J}_{\text{R}}(T_{\text{R}},T_{\text{L}}) + \mathcal{J}_{\text{R}}(T_{\text{L}},T_{\text{R}})}}{\text{Max}(\abs{\mathcal{J}_{\text{R}}(T_{\text{R}},T_{\text{L}})}, \abs{\mathcal{J}_{\text{R}}(T_{\text{L}},T_{\text{R}}))}},
	\end{equation}
	where the $\mathcal{J}_{\text{R}}(T_{\text{R}},T_{\text{L}})$ is the heat current from the right bath into the system for $T_{\text{R}} > T_{\text{L}}$, and vice versa  for  $\mathcal{J}_{\text{R}}(T_{\text{L}},T_{\text{R}})$ and $T_{\text{L}} > T_{\text{R}}$. 
	The rectification factor is a figure of merit that measures the quality of the diode, 
	taking
	values between 0 (symmetric heat flow) and 1 (perfect diode).
	
	In our numerical calculations we consider resonant 
	(RQ, $\omega_L = \omega_R$) 
	and off-resonant 
	(ORQ, $\omega_L \neq \omega_R$) 
	qubits for both Ohmic (OSD) and flat (FSD) spectral densities of the baths. 
	The
	simulations are done using scientific python packages along with 
	key libraries from QuTiP~\cite{qutip}. 
	\subsection{Effects of Asymmetric Exchange Interaction Strength}\label{hcAsymmetry}
	
	With ORQs, unit rectification is possible for wide range of system parameters. However, here we consider resonant case to show that unit rectifications is also possible for RQs. 
	Note that in thermal diode models proposed in~\cite{PhysRevE.89.062109,PhysRevE.94.042135,PhysRevE.95.022128}, rectification is not possible for RQs when each qubit is connected 
	to a separate
	thermal bath.
	
\begin{figure}[t]
		\centering
		\includegraphics[width=0.5\textwidth]{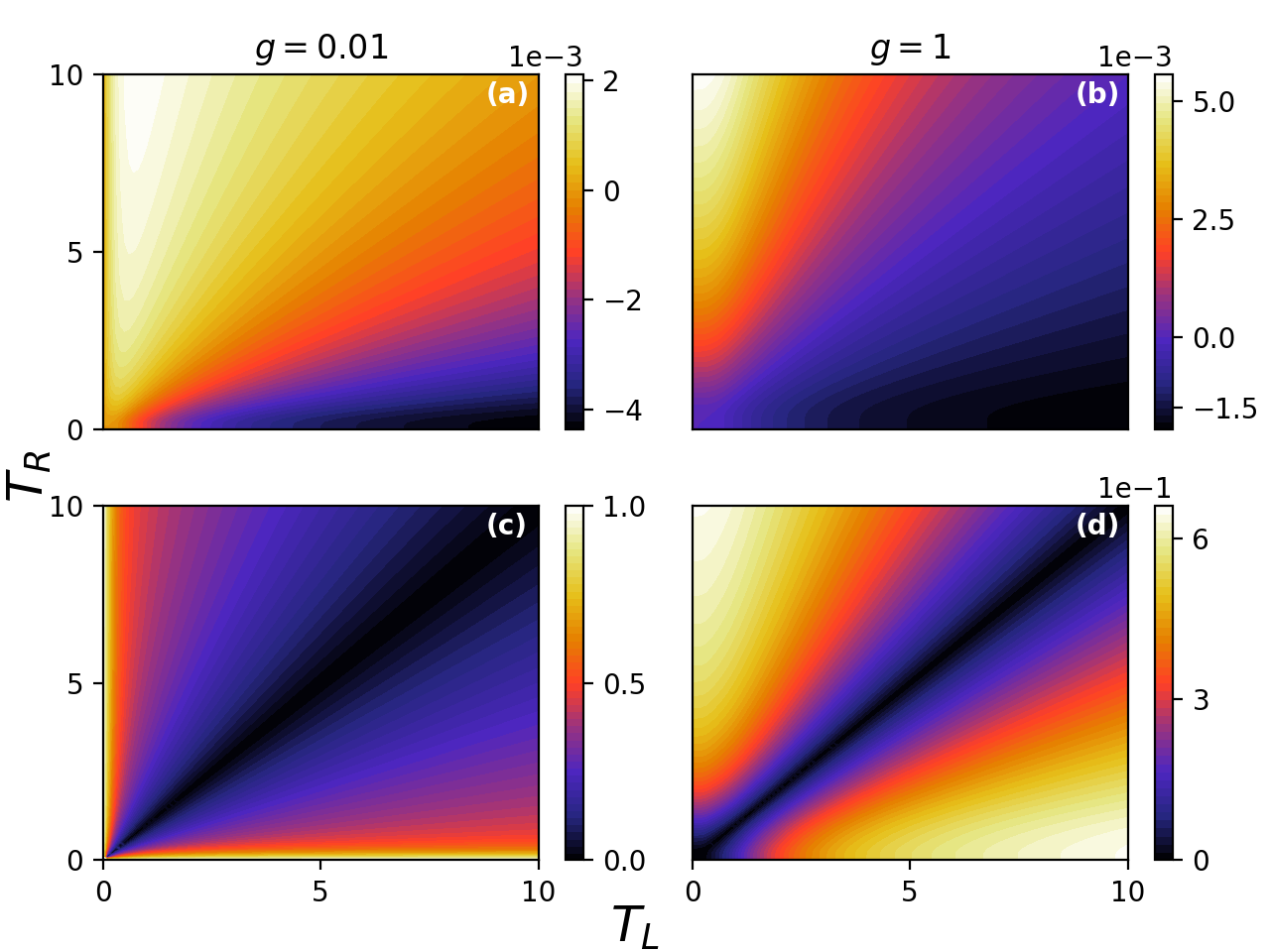}
		\caption{(Color online) Steady state heat current $\mathcal{J}_{\text{R}}$  and  rectification $\mathcal{R}$ as  functions of temperatures 
			$T_{\text{L}}$ and $T_{\text{R}}$, assuming heat baths with FSD. 
			The
			interaction strength is 
			$g=0.01$ in the left and $g=1$ in the right column.  Panels (a) and (b) show  $\mathcal{J}_{\text{R}}$, and panels (c) and (d) show the corresponding values of $\mathcal{R}$. The qubits are resonant with $\omega_{\text{L}} = \omega_{\text{R}}=1$, and the
			dissipation rates are $\kappa_{\text{LL}}=\kappa_{\text{RR}}=0.01$.
		}
		\label{fig:JRorqVSrq_Flat}
	\end{figure}

	In Fig.~\ref{fig:JRorqVSrq_Flat} we plot the heat current $\mathcal{J}_{\text{R}}$ 
	and the rectification factor $\mathcal{R}$
	for different values of $T_{\text{L}}$ and $T_{\text{R}}$ 
	with RQs and
	baths with FSD.
	As can be seen, the heat
	flow is 
	asymmetric with respect to
	the $T_{\text{L}}=T_{\text{R}}$ axis
	(Figs.~\ref{fig:JRorqVSrq_Flat}(a) and \ref{fig:JRorqVSrq_Flat}(b)):
	For 
	$T_{\text{R}}>T_{\text{L}}$, 
	the heat current
	$\abs{\mathcal{J}_{\text{R}}}$ is smaller 
	compared to the case when thermal biased is reversed so that the system rectifies 
	heat 
	from left to right.
	Although rectification is 
	stronger
	for off-resonant qubits than for resonant, 
	RQs bring crucial advantage 
	of much larger heat flows.
	For example, the  heat current  for RQs is two orders of magnitude larger in comparison to that of ORQs with $\omega_L = 20\omega_R$ and the same set of remaining parameters.	
	Temperature dependence of the rectification factor for two different values of the qubit-qubit coupling strength $g$ is shown
	in Fig.~\ref{fig:JRorqVSrq_Flat}(c) and \ref{fig:JRorqVSrq_Flat}(d). 
	Almost perfect diode behavior for $g = 0.01$ is obtained only with large temperature difference $\abs{T_{\text{R}}-T_{\text{L}}}$ 
	as shown in the Fig.~\ref{fig:JRorqVSrq_Flat}(c), and $60\%$ rectification is possible over wide range of bath temperatures if coupling strength is increased to $g = 1$ as shown in Fig.~\ref{fig:JRorqVSrq_Flat}(d).

	Our calculations show similar properties also for baths with OSD. However, the temperature domain of high rectification is then larger, while the heat current is reduced compared to the baths with FSD. This is related to the  asymmetry of the spectrum and the proportionality of the spectral response functions to the transition frequencies.	
	\subsection{Variation of Qubit Frequencies}\label{qubitVariation}
	
	
	
	\begin{figure}[t]
		\centering
		\includegraphics[width=0.5\textwidth]{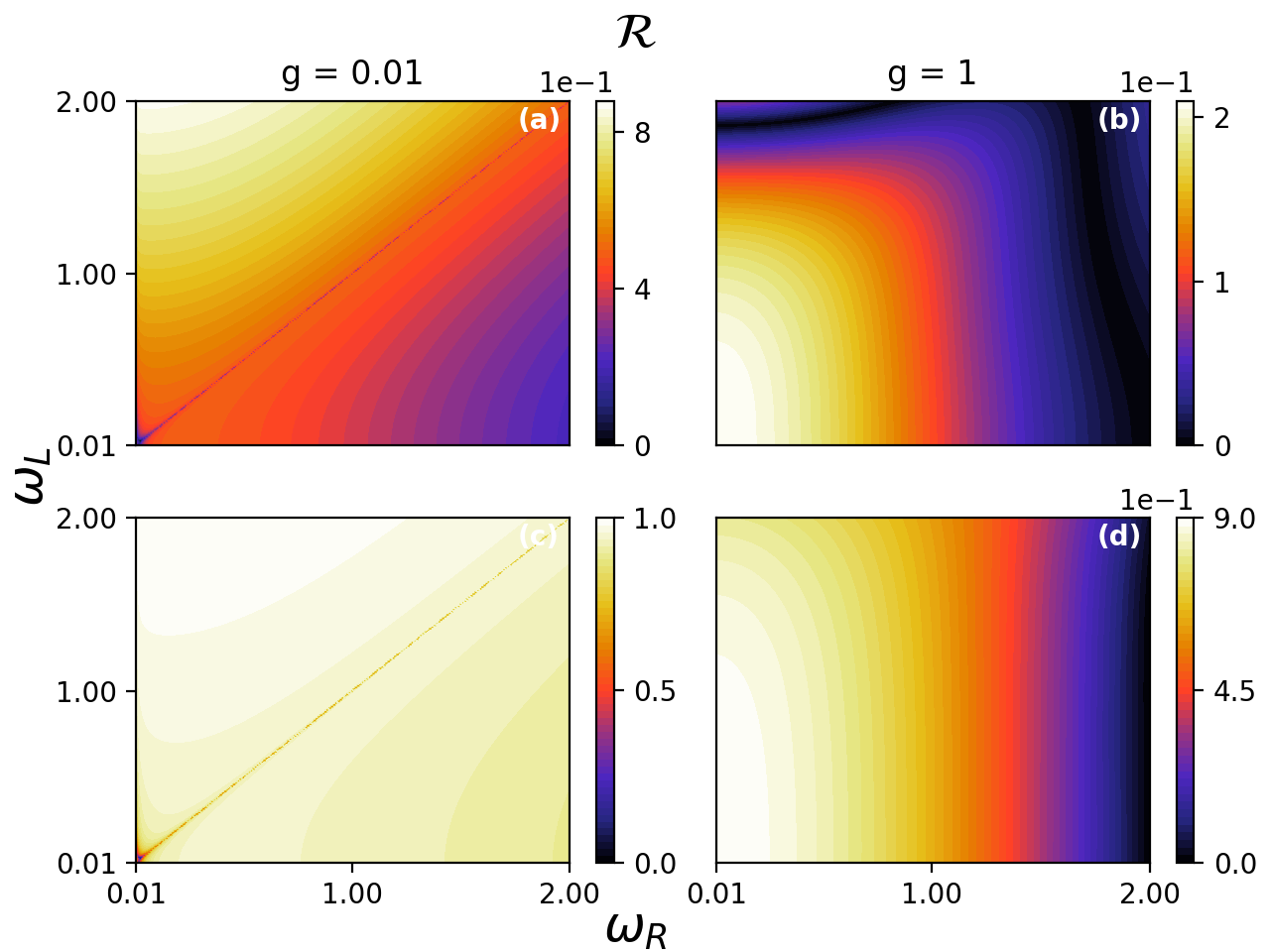}
		\caption{(Color online) Rectification $\mathcal{R}$ as a function of the qubit frequencies 
			$\omega_{\text{L}}$ and $\omega_{\text{R}}$.
			The interaction strength is 
			$g=0.5$ in the left and $g=1$ in the right column.  In panels (a) and (b) $T_{\text{R}} = 0.5T_{\text{L}} = 1$, and   in panels (c) and (d)  $T_{\text{R}} = 0.05T_{\text{L}} = 10$. 
			The remaining parameters are the same as in Fig.~\ref{fig:JRorqVSrq_Flat}.}
		\label{fig:flatHC}
	\end{figure} 
	

	In the previous subsection, we showed that the RQs behave as a thermal diode, and both quantitative and qualitative behaviors of rectification depends on coupling strength $g$. Other control parameters for rectification  in our model are tuning the qubit frequencies and/or qubit-bath coupling rates $\kappa_{\text{ij}}$, where $i,j = \text{L,R}$ baths.
	Let us now
	analyze the effect of variation of qubits frequencies on the rectification 
	magnitude.
	%
	%
	Fig.~\ref{fig:flatHC} shows 
	$\mathcal{R}$
	for two different coupling strengths, $g = 0.01$ in the left column and $g = 1$ in the right column. Figs.~\ref{fig:flatHC}(a) and \ref{fig:flatHC}(b) are for small temperature difference between the two baths, while Figs.~\ref{fig:flatHC}(c) and \ref{fig:flatHC}(d) 
	correspond to
	large temperature differences. 
	As can be seen,
	large temperature difference increases the diode quality irrespective of the strength of the qubit-qubit coupling. However, RQs 
	generate vanishing
	rectification for small value of coupling strength $g=0.01$ as depicted in Fig.~\ref{fig:flatHC}(c). 
	On the other hand,
	almost unit rectification of the heat current by RQs can be obtained if 
	the coupling strength 
	is increased to $g=1$ (Fig.~\ref{fig:flatHC}(d)). 
	All these results demonstrate that the diode quality can be controlled in a wide range of system parameters. 
	For instance, as shown in Fig.~\ref{fig:JRorqVSrq_Flat}(c), RQs model with $g = 0.01$ 
	behaves as a thermal diode
	for a narrow range of temperature values,
	however this region can be enhanced by changing the qubit frequencies as shown in Fig.~\ref{fig:flatHC}.
	System-bath coupling rates are another control parameter for the rectification, but here we consider them to be symmetric. A brief discussion to show the effect of dissipation rates on rectification is presented in the Appendix~\ref{AppendixB}.

	
\section{\label{sec:implementations}Physical Model Systems for Possible Implementations}
Asymmetric spin-spin interactions, such as $\sigma_L^{z}\sigma_R^{x}$, can be found in natural systems, for example, effectively in magnetic macromolecules in nuclear spin environments~\cite{tupitsyn_effective_1997} or directly in weak ferromagnetic systems with spin-orbit coupling~\cite{moriya_anisotropic_1960,dzyaloshinsky_thermodynamic_1958}. More recently similar spin bath models are used to explore
controlling dephasing of a single qubit~\cite{bhaktavatsala_rao_zeno_2011}. Our HD mechanism rely on such an interaction between two
qubits. We specifically use the same interaction ($\sigma_L^{z}\sigma_R^{x}$) proposed in Ref.~\cite{PhysRevA.92.062114}; where the model is motivated for its mathematical
similarity to optomechanical coupling but no physical motivation or an implementation scheme is given.
In order to produce such an interaction in a physical and controllable manner, we have suggested several systems in preceding sections. Here we will provide more details of these systems by presenting explicit model Hamiltonians.  
\subsection{Optomechanical route for $\sigma_L^{z}\sigma_R^{x}$}
The optomechanical coupling between an optical resonator of frequency $\omega$ and a mechanical resonator of frequency $\Omega$
can be written as~\cite{law_effective_1994,law_interaction_1995} ($\hbar =1$)
\begin{eqnarray}
H=\omega a^\dag a+\Omega b^\dag b + ga^\dag a(b+b^\dag),
\end{eqnarray}
where the annihilation and creation operators of photons and phonons are denoted by  $a,a^\dag$ and $b,b^\dag$, respectively. A statistical mutation
of bosonic operators to spin operators can be constructed as an inversion of spin to boson Holstein-Primakoff transformation such that~\cite{carneiro_application_2017}
\begin{eqnarray}
a^\dag a &=& J\mathds{1}+J_z,
\quad 
a=\frac{1}{\sqrt{J\mathds{1}+J_z}}J_-,\nonumber\\
a^\dag&=&J_+\frac{1}{\sqrt{J\mathds{1}+J_z}},
\end{eqnarray}
where $J$ is the total spin and $J_\pm,J_z$ satisfy the SU(2) algebra. Employing this transformation to both 
photons and phonons, and assuming weakly excited spins ($\langle J_z\rangle\ll J$) the bosonic optomechanical
model can be mapped to the asymmetric spin-spin coupling model. We remark that replacing bosonic
mechanical mode by a single spin-$1/2$ is employed in a circular quantum walk problem by assuming $\langle b^dag b\rangle\ll 1$ so that only the two lowest vibronic modes are accessible~\cite{Moqadam}.

The weak excitation condition as well as lack of physical
qubits make the optomechanical route is a limited and indirect approach to implement our HD scheme. The other restrictions
such as weak $g$ or frequency difference of optical and mechanical modes can be relaxed in electrical analogs of optomechanical-like
couplings~\cite{johansson_optomechanical-like_2014}.
\subsection{Coupled Raman model route for $\sigma_L^{z}\sigma_R^{x}$}
Hamiltonian of a system consisting of a three-level atom in a single mode cavity is given by  ($\hbar =1$)
\begin{eqnarray}
H&=&\omega a^\dag a +\omega_r|r\rangle\langle r|+\omega_e|e\rangle\langle e|+\omega_g|g\rangle\langle g|+\nonumber\\
&+&g(a^\dag(|g\rangle\langle r|+|e\rangle\langle r|)+\text{H.c.},
\end{eqnarray}
where $g$ is the cavity-atom coupling coefficient. $\omega_r,\omega_e,\omega_g$ denote the upper, middle, and lowest energy levels with the associated states 
$|r\rangle,|e\rangle,|g\rangle$, respectively. Assuming the lower doublet of energy levels are quasi-degenerate ($\omega_e\approx\omega_r$ 
and taking the 
detuning $\delta=\omega_r-\omega_e-\omega_a$ of the cavity mode from the atomic resonance  is much greater than $g$ then the upper level can
be adiabatically eliminated from the dynamics, which can be described by an effective Hamiltonian of the form
\begin{eqnarray}
H=\omega_c a^\dag a + \epsilon \sigma_z -\frac{g^2}{\delta}a^\dag a\sigma_x.
\end{eqnarray} 
This special two-photon transition model is known as the Raman coupled model~\cite{knight_quantum_1986,Phoenix:90}.
Here,  the intensity dependent Stark shift in the cavity frequency is neglected; we introduced $\epsilon=\omega_e-\omega_g$; and
$\sigma_z=|e\rangle\langle e|-|g\rangle\langle g|,\sigma_x=|e\rangle\langle g|+|g\rangle\langle e|$.
Similar to the optomechanical case, one can assume weak excitation of the cavity mode and replace $a^\dag a$ with $\sigma_L^{z}$
to get the desired asymmetric spin-spin interaction effectively. 

We remark that the Raman route is quite generic provided that we assume only
two vibronic levels are involved in an optical Raman scattering from a molecule. The Raman scattering of a field $E$ is described by the interaction
$q|E|^2$ where $q$ is the vibrational displacement which can be replaced by $\sigma_x$ in the case of two-level approximation for the vibrational
motion. Using $E\sim (a+a^\dag)$ and neglecting two photon terms, $a^\dag a \sigma_x$ is obtained~\cite{schoendorff_analytic_1990}. 
\subsection{Quantum walk with trapped ions scheme for $\sigma_L^{z}\sigma_R^{x}$}
The quantum walk on a circle can be simulated with trapped ions where the steps of the walker are taken in quantum optical phase space
according to the single step generator~\cite{PhysRevA.65.032310}
\begin{eqnarray}
U=\text{e}^{ip\sigma_z}H,
\end{eqnarray}
where $p$ is the momentum operator generating the step conditioned by the result of the 
coin toss operation. Here $H$ stands for the Hadamard gate operator. It is proposed that the step can be implemented using
four Raman beam pulses sequentially~\cite{PhysRevA.65.032310}. Assuming steps are small such that vibrational excitation is much less than $1$ then
again the effective Hamiltonian associated with the step generator corresponds to asymmetric spin-spin interaction $\sigma_L^{z}\sigma_R^{x}$.
\subsection{Circuit QED scheme for $\sigma_L^{z}\sigma_R^{x}$}
General Hamiltonian of a superconducting resonator interacting with superconducting qubit can be expressed as~\cite{blais_cavity_2004}
\begin{eqnarray}
H=\omega a^\dag a+\frac{\Omega}{2}+g(\cos{\theta}\sigma_z+\sin{\theta}\sigma_x)(a+a^\dag),
\end{eqnarray}
where $\omega$ is the frequency of the resonator with the annihilation and creation operators $a$ and $^\dag$, respectively.
The qubit frequency is denoted by $\Omega$. The coupling coefficient $g$ and the mixing angle $\theta$ depend on Josephson-Junction
properties. By adjusting the junction parameters to get $\theta=0$ one gets the so called phase-gate term~\cite{blais_quantum-information_2007}, 
which becomes 
$\sigma_L^{z}\sigma_R^{x}$ if the resonator
is weakly excited.
\subsection{Two-qubit Raman coupled scheme for $\sigma_L^{z}\sigma_R^{x}$}
\begin{figure}[t]
		\centering
		\includegraphics[width=8 cm]{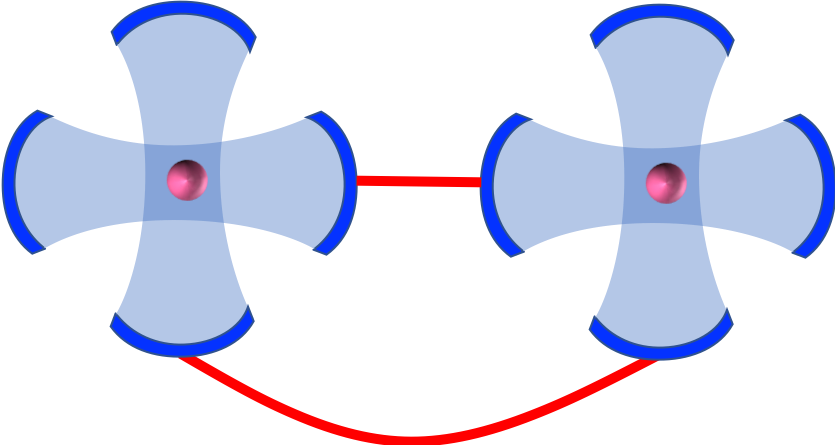}
		\caption{(Color online) Schematic diagram of two distant three-level atoms trapped inside fiber coupled bi-modal cavities.}
		\label{fig:implementation}
	\end{figure}

\begin{figure}[t]
		\centering
		\includegraphics[width=8.3 cm]{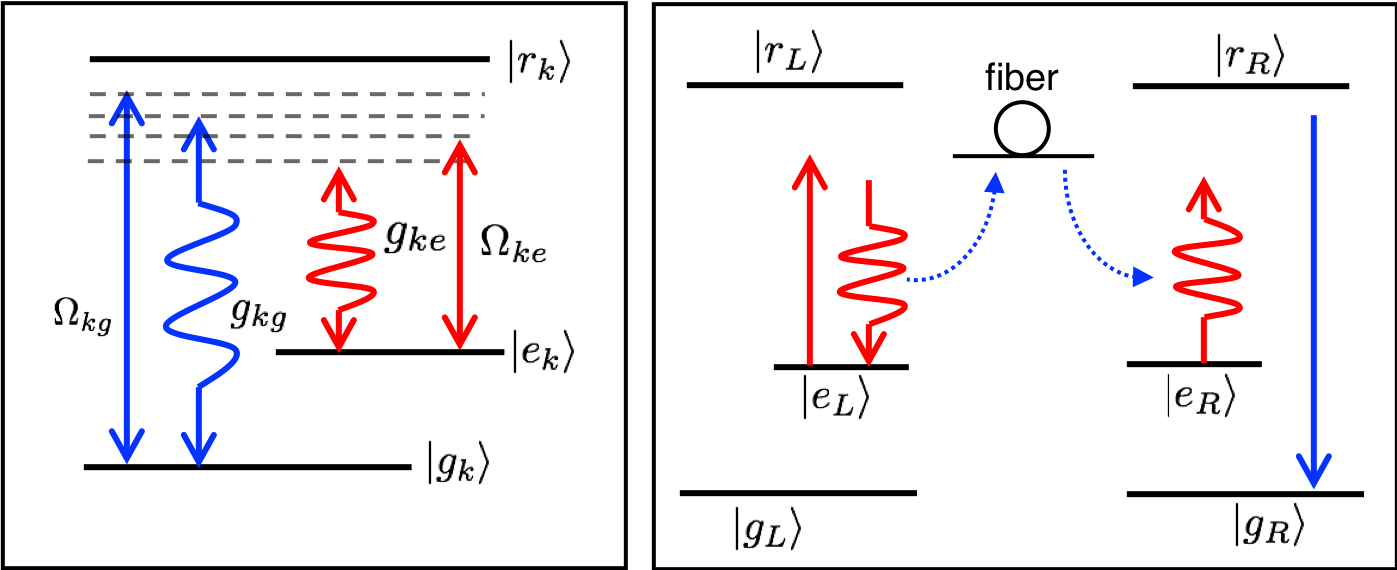}
		\caption{(Color online) (Left panel) The classical laser and cavity field driven transitions in the 
		three-level atoms, distinguished by label $x=L,R$ for the atom trapped in the left or right cavity (cf.~Fig.~\ref{fig:implementation}). 
		Both the classical and cavity fields drive the transitions. Upper, middle, and lower atomic states are denoted by $|r_x\rangle,|e_x\rangle$,
		and $g_x\rangle$, respectively. (Right panel) Special case where only the $|r_L\rangle\leftrightarrow |e_L\rangle$ is driven by the both 
		classical laser and the cavity field. The other fields on the left atom are turned off. The right atom is driven by a single classical and a single 
		cavity field in Raman configuration.}
		\label{fig:specialImplementation}
	\end{figure}

So far we have considered effective qubit systems to implement $\sigma_L^{z}\sigma_R^{x}$ coupling. Let us now assume a pair of three level
atoms, each held separately in two bi-modal optical cavities. The cavities are coupled to each other via single-mode fibers as depicted
in Fig.~\ref{fig:implementation}. This scheme is a generalization of the one for single-mode cavities 
described in Ref.~\cite{zheng_arbitrary_2010}. We consider the general case where the
atomic transitions are driven by both classical laser and cavity fields. The interaction picture Hamiltonian describing the coupling of 
atoms and lasers is ($\hbar=1$)
\begin{eqnarray}
H_{1}=\sum_{k=L,R}\sum_{x=e,g}\Omega_{kx}\text{e}^{i\Delta_{kx}t}|r_k\rangle\langle x_k|+\text{H.c.},
\end{eqnarray}
where $\Omega_{kx}$ is the Rabi frequency associated with the transition $|r_k\rangle\leftrightarrow |x_k\rangle$ and $\Delta_{kx}$ denotes the  detuning of the laser from the respective transition. Similarly, interaction of the cavity modes and the atoms are expressed in the interaction picture
as
\begin{eqnarray}
H_{2}=\sum_{k=L,R}\sum_{x=e,g}g_{kx}a_{kx}\text{e}^{i\delta_{x}t}|r_k\rangle\langle x_k|+\text{H.c.}.
\end{eqnarray}
Here, $a_{kx}$ and $a_{kx}^\dag$ are the annihilation and creation operators for the respective cavity modes and $\delta_{x}$ is the cavity
detuning from the respective transition. The cavities are assumed to be connected to each other through single-mode (short) fibers. The optical
interactions are described by the Hamiltonian
\begin{eqnarray}
H_{3}=\sum_{e,g}\nu_{x}b_{x}(a_{Lx}^\dag+a_{Rx}^\dag)+\text{H.c.},
\end{eqnarray}
where $g_{kx}$ is the atom-cavity coupling coefficient, 
$\nu_x$ is the coupling strength of the fiber mode and the respective cavity mode. Creation and annihilation operators for the fiber
modes are denoted by $b_x$ and $b_x^\dag$, respectively. The notation is shown in Fig.~\ref{fig:specialImplementation} for clarity. We remark that these Hamiltonians are straightforward direct generalization of
those in Ref.~\cite{zheng_arbitrary_2010}. The only new ingredient here is that we allow for an additional classical drive and a cavity field
acting on each atom such that each transition can be driven by both classical and cavity fields. 

The full Raman coupled model allows us
to engineer a variety of qubit-qubit interactions, including asymmetric ones. In particular we can generalize the example given in
Ref.~\cite{zheng_arbitrary_2010}, where it is shown that if both the classical and cavity fields are driving only the $|r\rangle\leftrightarrow |g\rangle$
transition then the qubit-qubit interaction would of the form $\sim |g_L\rangle\langle g_L|\otimes |g_R\rangle\langle g_R|$. In our general
case we can assume the similar situation for only one qubit such that  $|r_L\rangle\leftrightarrow |g_L\rangle$ is driven by both the classical and
cavity fields. While in the second qubit, we assume classical field drives the $|r_R\rangle\leftrightarrow |g_R\rangle$ and cavity field drives
the $|r_R\rangle\leftrightarrow |e_R\rangle$ transition. The scheme is depicted in Fig.~\ref{fig:specialImplementation}. Accordingly, this would yield an effective qubit-qubit coupling $|e_L\rangle\langle e_L|\otimes
|g_R\rangle\langle e_R|$
that depends on 
the population of the $|e_L\rangle$. Taking into account the Hermitian conjugate process and expressing the coupling in terms of the Pauli spin
operators we get the $\sigma_L^{z}\sigma_R^{x}$ interaction.

	\section{\label{sec:Conclusion}Conclusion}
	We 
	have
	investigated a quantum thermal diode composed of two qubits coupled to each other via anisotropic exchange interaction originally introduced in Ref.~\cite{PhysRevA.92.062114}.
	By deriving the 
	global master equation, analytical expressions for the heat currents in the system 
	are found.
	We have used the rectification factor to quantify the diode quality, calculating the results for both FSD and OSD of the thermal reservoirs.
	The rectification mechanism 
	is explained in terms of 
	the anisotropic exchange interactions: the baths can excite only some of the transitions in the Raman or four-wave-mixing global cycles so that some global cycles can run only in one direction and not oppositely.
	We have shown that in our model the diode behavior relies neither on the asymmetry of qubit frequencies nor
	on the difference of dissipation rates of the heat baths. Rectification can be achieved even for resonant qubits thus allowing to conduct large heat currents without compromising rectification efficiency. Since the anisotropic exchange model  can be applied to natural weak ferromagnets~\cite{moriya_anisotropic_1960,dzyaloshinsky_thermodynamic_1958}, nuclear spin environments~\cite{bhaktavatsala_rao_zeno_2011,tupitsyn_effective_1997}, cavity QED~\cite{knight_quantum_1986,schoendorff_analytic_1990,Phoenix:90}, circuit QED~\cite{johansson_optomechanical-like_2014,blais_cavity_2004,blais_quantum-information_2007,zheng_arbitrary_2010}, trapped ions~\cite{PhysRevA.65.032310}, optomechanical systems~\cite{PhysRevA.92.062114,Moqadam,johansson_optomechanical-like_2014} we anticipate that our results can be significant for the heat management in such systems.
	\section{acknowledgments}
	\"{O}.~E.~M acknowledges fruitful discussions with W.~Niedenzu and R.~Uzdin. C.~K.~thanks to hospitality of QUB-CTAMOP and Reykjavik University Nanophysics groups where some part of this work had been completed. C.~K.~ thanks to M.~Paternostro, A.~Manolescu, and N. Aral for useful discussions. \"{O}.~E.~M and C.~K.~acknowledges support from the Ko\c{c} University T\"{U}PRA\c{S} Energy Center (KUTEM). T.O. acknowledges support of the Czech Science Foundation (GA\v{C}R), grant 17-20479S. G.K acknowledges the support of DFG, ISF and SAERI.
	\appendix
	\section{\label{AppendixME} Global Master equation}
	We present the derivation of global master equation given in Eq.~(\ref{eq:master}). The system Hamiltonian is diagonalized using the unitary transformation
	\begin{equation}\label{eq:transformation}
	U := \exp\Big(-i\frac{\theta}{2}\hat{\sigma}_{\text{L}}^{z}\hat{\sigma}_{\text{R}}^{y}\Big),
	\end{equation}
	where the angle $\theta$ is defined as
	\begin{eqnarray}
	\sin\theta := \frac{2g}{\Omega}, \qquad
	\cos\theta := \frac{\omega_\text{R}}{\Omega},\qquad
	\tan\theta := \frac{2g}{\omega_\text{R}},
	\end{eqnarray}
	such that $\Omega := \sqrt{\omega^2_\text{R} + 4 g^2}$. The transformed operators then read
	\begin{eqnarray}
	\hat{\tilde{\sigma}}_{\text{L}}^{x}&=& U \hat{\sigma}_{\text{L}}^{x}U^\dagger= \cos\theta\hat{\sigma}_{\text{L}}^{x} + \sin\theta\hat{\sigma}_{\text{L}}^{y}\hat{\sigma}_{\text{R}}^{y},\label{eq:transform}\\
	\hat{\tilde{\sigma}}_{\text{L}}^{y}&=& U \hat{\sigma}_{\text{L}}^{y}U^\dagger= \cos\theta\hat{\sigma}_{\text{L}}^{y} - \sin\theta\hat{\sigma}_{\text{L}}^{x}\hat{\sigma}_{\text{R}}^{y},\\
	\hat{\tilde{\sigma}}_{\text{L}}^{z}&=& U \hat{\sigma}_{\text{L}}^{z}U^\dagger= {\sigma}_{\text{L}}^{z},
	\end{eqnarray}
	and
	\begin{eqnarray}
	\hat{\tilde{\sigma}}_{\text{R}}^{x}&=& U \hat{\sigma}_{\text{R}}^{x}U^\dagger= \cos\theta\hat{\sigma}_{\text{R}}^{x} - \sin\theta\hat{\sigma}_{\text{L}}^{z}\hat{\sigma}_{\text{R}}^{z},\label{eq:transformm}\\
	\hat{\tilde{\sigma}}_{\text{R}}^{y}&=& U \hat{\sigma}_{\text{R}}^{y}U^\dagger= {\sigma}_{\text{R}}^{y},\\
	\hat{\tilde{\sigma}}_{\text{R}}^{z}&=& U \hat{\sigma}_{\text{R}}^{z}U^\dagger= \cos\theta\hat{\sigma}_{\text{R}}^{z} - \sin\theta\hat{\sigma}_{\text{L}}^{z}\hat{\sigma}_{\text{R}}^{x}.\label{eq:transforme}
	\end{eqnarray}
	The back transformations from dressed operators to bare operators reads from the Eqs.~(\ref{eq:transform})-(\ref{eq:transforme}) by switching dressed operators to bare operators and vice versa with $\theta$ replaced by $-\theta$.
	Then, with the transformation Eq.~(\ref{eq:transformation}), the Hamiltonian in Eq.~(\ref{eq:HamiltonianBare}) is diagonalized to the Hamiltonian given in Eq.~(\ref{eq:diaghamil}).  
	
	Eigenstates of the dressed Hamiltonian are given by the individual eigenstates of the qubits as,
	\begin{eqnarray}\label{eq:eigenstate1}
	\ket{1} = \cos\frac{\theta}{2}\ket{++} - \sin\frac{\theta}{2}\ket{+-} \\
	\ket{2} = \sin\frac{\theta}{2}\ket{++} + \cos\frac{\theta}{2}\ket{+-} \\
	\ket{3} = \cos\frac{\theta}{2}\ket{-+} + \sin\frac{\theta}{2}\ket{--} \\
	\ket{4} = \cos\frac{\theta}{2}\ket{--} - \sin\frac{\theta}{2}\ket{-+}\label{eq:eigenstate4},
	\end{eqnarray}
	with their corresponding eigenvalues $\omega_{1} = \frac{1}{2}(\omega_{\text{L}} + \Omega)$,  $\omega_{2} = \frac{1}{2}(\omega_{\text{L}} - \Omega)$, $\omega_{3} = \frac{1}{2}(-\omega_{\text{L}} + \Omega)$, and $\omega_{4} = \frac{1}{2}(-\omega_{\text{L}} - \Omega)$, respectively.
	
	The qubits are coupled to two baths of temperature $T_{\text{R}}$ and $T_{\text{L}}$ via the the Hamiltonian $\hat{H}_{SB}^{ij} = \hat{\sigma}_{i}^{x}\otimes\sum_{k}g_{k}^{j}(\hat{a}_{k}^{j} + \hat{a}_{k}^{j\dagger} )$,
	where $g_{k}^{j}$ are the coupling strengths to baths, and $\hat{a}_{k}^{j}$ ($\hat{a}_{k}^{j\dagger}$) are the creation (annihilation) operator of the $k$ mode of the bath $i,j = \text{L,R}$, whose Hamiltonian is $\hat{H}_{j} = \sum_{k}\omega_{k}\hat{a}_{k}^{j\dagger}\hat{a}_{k}^{j}$. To calculate the master equation, we move to the interaction picture in which
	\begin{eqnarray}\label{eq:tdependeqs}
	\hat{\sigma}_{\text{L}}^{x}(t)&=&\cos\theta\hat{\tilde{\sigma}}_{\text{L}}^{-}e^{-i\omega_\text{L}t}-\sin\theta\hat{\tilde{\sigma}}_{\text{L}}^{+}\hat{\tilde{\sigma}}_{\text{R}}^{-}e^{-i(\Omega-\omega_\text{L})t} \\ \nonumber&&+ \sin\theta\hat{\tilde{\sigma}}_{\text{L}}^{-}\hat{\tilde{\sigma}}_{\text{R}}^{-}e^{-i(\Omega+\omega_\text{L})t} + \mathrm{H.c.}\\
	\hat{\sigma}_{\text{R}}^{x}(t)&=& \cos\theta\hat{\tilde{\sigma}}_{\text{R}}^{-}e^{-i\Omega t} + \frac{1}{2}\sin\theta\hat{\tilde{\sigma}}_{\text{L}}^{z}\hat{\tilde{\sigma}}_{\text{R}}^{z} + \mathrm{H.c.}
	\end{eqnarray}
	Hence, the master equation in the interaction picture is found to be the one that is given in Eq.~(\ref{eq:master}), with 
	\begin{eqnarray}
	G_{ij}(\omega)=
	\begin{cases}
	\kappa_{ij}(\omega)[1 + \bar{n}_{j}(\omega)], & \omega > 0 \\
	\kappa_{ij}(\abs{\omega})\bar{n}_{j}(\abs{\omega}), & \omega < 0 \\
	0, & \omega = 0
	\end{cases},
	\end{eqnarray}
	where $\bar{n}_{j}(\omega) := 1/(\exp{\omega/T_{j}} - 1)$ is the average excitation number of $j = \text{L,R}$ baths. We define $\kappa_{ij}(\omega)$ as rates, which are independent of frequency for the flat spectrum, and $\kappa_{ij}(\omega) = \kappa_{ij}\omega$ for the Ohmic spectrum.
	
	\begin{eqnarray}
	\label{eq:L_L}
	\hat{ \mathcal{L}}_{\text{Lj}} &=& G_{\text{Lj}}(\omega_{\text{L}})\cos^{2}\theta\hat{\mathcal{D}}[\hat{\tilde{\sigma}}_{\text{L}}^{-}]
	+ G_{\text{Lj}}(-\omega_{\text{L}})\cos^{2}\theta\hat{\mathcal{D}}[\hat{\tilde{\sigma}}_{\text{L}}^{+}] \\ \nonumber&+& G_{\text{Lj}}(\omega_{\text{23}})\sin^{2}\theta\hat{\mathcal{D}}[\hat{\tilde{\sigma}}_{\text{L}}^{-}\hat{\tilde{\sigma}}_{\text{R}}^{+}]
	+ G_{\text{Lj}}(-\omega_{\text{23}})\sin^{2}\theta\hat{\mathcal{D}}[\hat{\tilde{\sigma}}_{\text{L}}^{+}\hat{\tilde{\sigma}}_{\text{R}}^{-}]
	\\ \nonumber &+& G_{\text{Lj}}(\omega_{\text{14}})\sin^{2}\theta\hat{\mathcal{D}}[\hat{\tilde{\sigma}}_{\text{L}}^{-}\hat{\tilde{\sigma}}_{\text{R}}^{-}]
	+ G_{\text{Lj}}(-\omega_{\text{14}})\sin^{2}\theta\hat{\mathcal{D}}[\hat{\tilde{\sigma}}_{\text{L}}^{+}\hat{\tilde{\sigma}}_{\text{R}}^{+}],
	\\
	\hat{ \mathcal{L}}_{\text{Rj}}&=& G_{\text{Rj}}(\Omega)\cos^{2}\theta\hat{\mathcal{D}}[\hat{\tilde{\sigma}}_{\text{R}}^{-}]
	+ G_{\text{Rj}}(-\Omega)\cos^{2}\theta\hat{\mathcal{D}}[\hat{\tilde{\sigma}}_{\text{R}}^{+}], \label{eq:L_R},
	\end{eqnarray}
	where the first index represents left (L) or right (R) qubit, and the second index $j = \text{L,R}$ is for baths with $\theta = \arctan(2g/\omega_{\text{R}})$.  In the main text, we consider $\kappa_{ij} = 0$ for distinct $i$ and $j$. The Lindblad dissipators $\hat{\mathcal{D}}[\hat{A}]$ are defined as~\cite{lind1,lind2}
	\begin{equation}\label{dissipator}
	\hat{\mathcal{D}}[\hat{A}] = \hat{A}\hat{\rho}\hat{A}^{\dagger} - \frac{1}{2}\left(\hat{A}^{\dagger}\hat{A}\hat{\rho} + \hat{\rho}\hat{A}^{\dagger}\hat{A}\right).
	\end{equation}
	
	\section{\label{AppendixB} Case of Spatially Overlapping Thermal Baths Over the Both Qubits}
	\label{app:}
		
	\begin{figure}[t]
		\centering
		\includegraphics[width=0.5\textwidth]{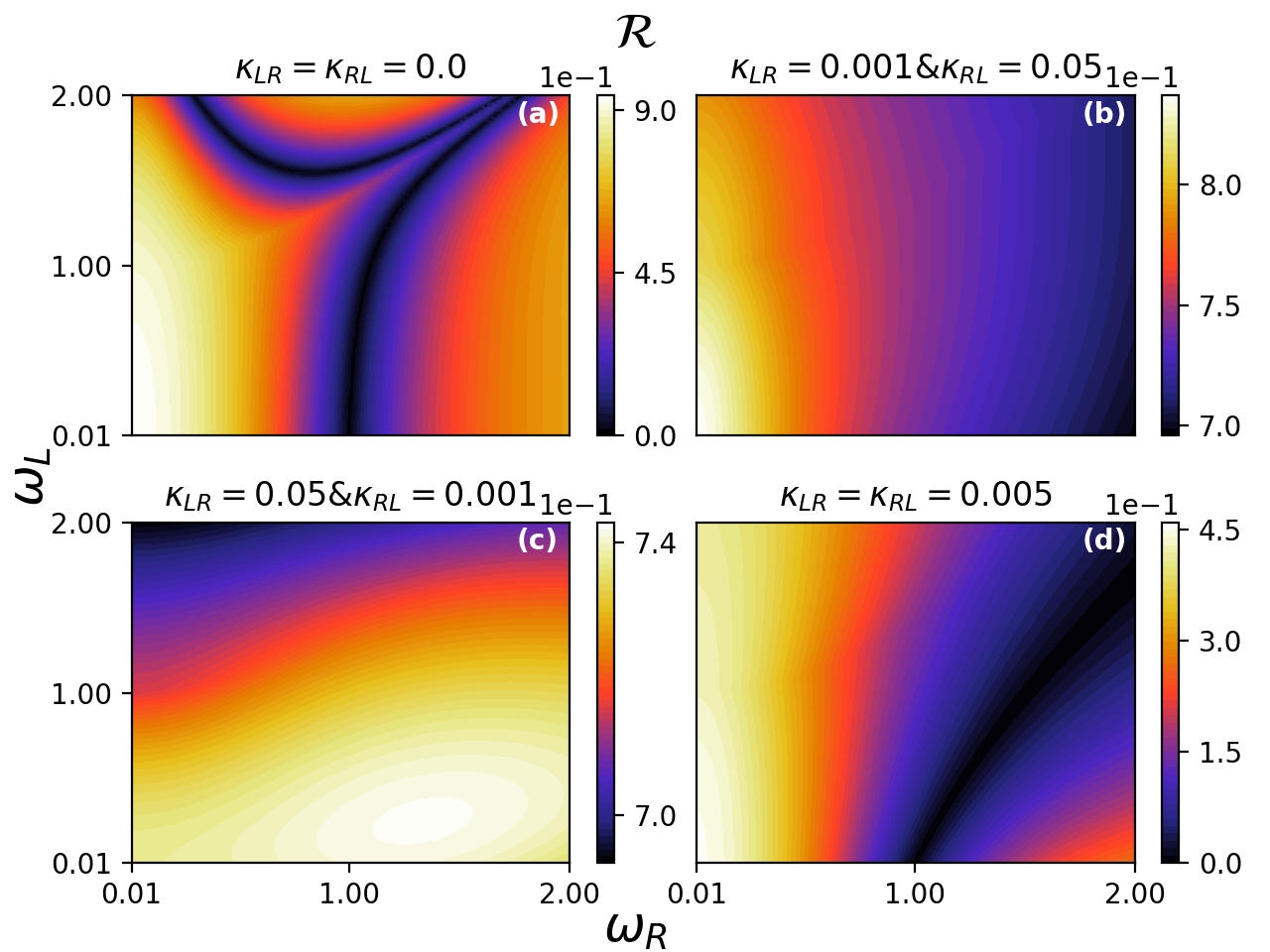}
		\caption{(Color online) Rectification $\mathcal{R}$ for the cases with different coupling rates between both baths to both qubits. Panel (a) shows the local bath case. Panels (b) and (c) show the case with both baths accessing their local qubits symmetrically but asymmetrically to the other one. Panel (c) is the case with all the qubit-bath couplings being symmetric. In all panels, $T_\text{L}=10$, $T_\text{R}=0.5$, and $\kappa_\text{LL}=\kappa_\text{RR}=0.01$. Rest of the parameters are the same as in Fig.~\ref{fig:JRorqVSrq_Flat}.}
		\label{fig:ghigh}
	\end{figure}
	In our treatment, we have assumed that each bath is only connected to its corresponding qubit. 
	The assumption of local baths for the qubits however is
	restrictive for possible embodiments of our thermal diode scheme. In practice, thermal baths can overlap spatially over the closely spaced pair of interacting qubits. Special systems where such an overlap can be exactly absent could still be found. One scenario is to use optomechanical like coupling of two transmission line resonators as proposed in Ref.~\cite{johansson_optomechanical-like_2014}. If the resonators are weakly excited to the limit of a single photon, then the system mimics our case of two interacting qubits with local baths realized by using thermal noise currents fed into the transmission line resonators. For other implementations, such as using trapped ions~\cite{PhysRevA.65.032310}, off-resonant Raman systems~\cite{Phoenix:90}, or asymmetric exchange interactions~\cite{moriya_anisotropic_1960,dzyaloshinsky_thermodynamic_1958}, spatial overlap of the thermal baths can be unavoidable without taking additional measures such as using local modulation of the qubits by external drives to effectively make the
	baths local~\cite{gordon_preventing_2006,gordon_scalability_2011}. Apart from such extra design complications, we would like to now address the question to which extent spatial overlap of the baths over the qubit pair can be tolerated for a significant thermal rectification. 
	For that aim, we consider a generalization of our master equation Eq.~\ref{eq:master} by including the cross terms $\hat{ \mathcal{L}}_{\text{LR}}$ and $\hat{ \mathcal{L}}_{\text{RL}}$ describing the access of the non-local thermal baths to both qubits in the bare state picture.
	
	Fig.~\ref{fig:ghigh}(a) shows the case $\kappa_\text{LR}=\kappa_\text{RL}=0$, which is equivalent to local case given in Eq.~\ref{eq:master}. 
	We have considered two cases, where the spatial overlap of the baths may lead to the same or different coupling rates to the distant qubits.
	In all cases, spatial overlap degrades the thermal diode quality by decreasing the maximum rectification factor. On the other hand, in the case of different coupling rates of the baths to their distant qubits we have found that the lowest value of the rectification factor is increased,
	as shown in Figs.~\ref{fig:ghigh}(b) and \ref{fig:ghigh}(c). The decrease of the diode quality is severe in the case of same coupling rates of the baths to the distant qubits. When the cross-coupling rate is about half of the local coupling rate then the diode operates at half efficiency,
	where the rectification factor is reduced from $\sim 90\%$ to  $\sim 45\%$. When all the coupling rates are equal to each other the rectification is practically lost.
	
	We conclude that if the spatial overlap of the thermal baths cannot be avoided then either asymmetric coupling rates to distant qubits 
	should be sought for or local modulators on qubits should be implemented~\cite{gordon_preventing_2006,gordon_scalability_2011}.
	
	\section{\label{AppendixC} Dynamics and the Heat Currents}
	\label{app:dynamics}
	Equations of motions for the relevant dynamical observables of our system are 
	determined from the master equation, Eq.~(\ref{eq:master}), and given by
	\begin{eqnarray}\label{eq:dynamics}
	\dfrac{d}{dt}\langle\hat{\tilde{\sigma}}^z_\text{L}\rangle &=& \cos^{2}\theta [G_\text{L}(-\omega_\text{L})\langle\mathds{A}\rangle -G_\text{L}(\omega_\text{L})\langle\mathds{B}\rangle] \\ \nonumber&&  + \frac{1}{2}\sin^{2}\theta[ G_\text{L}(\omega_\text{1})\langle\mathds{A}\mathds{D}\rangle-G_\text{L}(-\omega_\text{1})\langle\mathds{B}\mathds{C}\rangle \\ \nonumber && + G_\text{L}(-\omega_\text{2})\langle\mathds{A}\mathds{C}\rangle - G_\text{L}(\omega_\text{2})\langle\mathds{B}\mathds{D}\rangle] \\
	\dfrac{d}{dt}\langle\hat{\tilde{\sigma}}^z_\text{R}\rangle &=&\cos^{2}\theta [G_\text{R}(-\Omega)\langle\mathds{C}\rangle -G_\text{R}(\Omega)\langle\mathds{D}\rangle] \\ \nonumber&&+ \frac{1}{2}\sin^{2}\theta[ -G_\text{L}(\omega_\text{1})\langle\mathds{A}\mathds{D}\rangle+G_\text{L}(-\omega_\text{1})\langle\mathds{B}\mathds{C}\rangle  \\ \nonumber&&+ G_\text{L}(-\omega_\text{2})\langle\mathds{A}\mathds{C}\rangle - G_\text{L}(\omega_\text{2})\langle\mathds{B}\mathds{D}\rangle] 
	\end{eqnarray}
	\begin{eqnarray}
	\nonumber\dfrac{d}{dt} \langle\hat{\tilde{\sigma}}^z_\text{L}\hat{\tilde{\sigma}}^z_\text{R}\rangle &=& \cos^{2}\theta[G_\text{L}(-\omega_{\text{L}})\langle\mathds{A}\hat{\tilde{\sigma}}^z_\text{R}\rangle - G_\text{L}(\omega_{\text{L}})\langle\mathds{B}\hat{\tilde{\sigma}}^z_\text{R}\rangle \\ && + G_\text{R}(-\Omega)\langle\mathds{D}\hat{\tilde{\sigma}}^z_\text{L}\rangle - G_\text{R}(\Omega)\langle\mathds{C}\hat{\tilde{\sigma}}^z_\text{L}\rangle ], 
	\end{eqnarray}
	where we introduce the operators $\mathds{A} = (\mathds{1} - \hat{\tilde{\sigma}}_{\text{L}}^{z})$, $\mathds{B} = (\mathds{1} + \hat{\tilde{\sigma}}_{\text{L}}^{z})$, $\mathds{C} = (\mathds{1} - \hat{\tilde{\sigma}}_{\text{R}}^{z})$, and $\mathds{D} = (\mathds{1} + \hat{\tilde{\sigma}}_{\text{R}}^{z})$. 
	
	The heat currents evaluate to
	\begin{eqnarray}
	\mathcal{J}_\text{L} &=& \frac{1}{2}\omega_\text{L}\cos^{2}\theta [G_\text{L}(-\omega_\text{L})\langle\mathds{A}\rangle -G_\text{L}(\omega_\text{L})\langle\mathds{B}\rangle] \\  \nonumber&+& \frac{1}{4}\sin^{2}\theta[\omega_{\text{1}}G_\text{L}(-\omega_{\text{1}})\langle\mathds{B}\mathds{C}\rangle  - \omega_{\text{1}}G_\text{L}(\omega_{\text{1}})\langle\mathds{A}\mathds{D}\rangle \\ &-& \omega_{\text{2}}G_\text{L}(\omega_{\text{2}})\langle\mathds{B}\mathds{D}\rangle + \omega_{\text{2}} G_\text{L}(-\omega_{\text{2}})\langle\mathds{A}\mathds{C}\rangle],\nonumber \\
	\mathcal{J}_\text{R} &=& \frac{1}{2}\Omega\cos^{2}\theta[G_\text{R}(-\Omega)\langle\mathds{C}\rangle -G_\text{R}(\Omega)\langle\mathds{D}\rangle]\label{hc}.\label{eq:appheat}
	\end{eqnarray}
	Substitution of transformed operators relation given in Eq.~(\ref{eq:transforme}) to Eq.~(\ref{hc}) yields the heat current relation in terms of bare operators and given in Eq.~(\ref{rightHC}). Even though, the Eqs.~(\ref{eq:dynamics})-(\ref{eq:appheat}) are easy to solve, the analytical expressions are too cumbersome to report her. However, we confirmed  that $\mathcal{J}_\text{L} + \mathcal{J}_\text{R} = 0$ at the steady state.
	%
\end{document}